\documentclass[aps, 
pra, 
superscriptaddress, reprint, 
longbibliography]{revtex4-2}
\usepackage{mathtools}
\usepackage{amsmath,amssymb,bm,bbm}
\usepackage{amsthm}
\usepackage{dsfont}
\usepackage{bbold}
\usepackage{bm}
\usepackage{braket}
\usepackage{csquotes}
\usepackage{comment}
\usepackage{float}
\usepackage{graphicx}
\usepackage{tikz}
\usepackage{xcolor}
\usepackage{subcaption}
\usepackage[section]{placeins}
\usepackage{ragged2e}
\usepackage{orcidlink}
\usepackage[capitalize]{cleveref}

\usepackage{hyperref}
\newtheorem{proposition}{Proposition}
\newtheorem{lemma}{Lemma}

\DeclareMathOperator{\Var}{Var}
\DeclareMathOperator{\Tr}{Tr}
\newcommand{\id}{\mathbbm 1}

\DeclareMathOperator{\supp}{supp}

\begin{document}
\preprint{APS/123-QED}
\title{Non-Hermitian entropy production from fluctuation theorems}
\author{Vasco Cavina\,\orcidlink{0000-0001-9468-7410}}
\email[Email: ]{vasco.cavina@sns.it}
\affiliation{Scuola Normale Superiore, 56126 Pisa, Italy}
\author{Frank Ernesto Quintela Rodr\'iguez\,\orcidlink{0000-0002-9475-2267}}
\email[Email: ]{frank.quintela@uam.es}
 \affiliation{Departamento de Física Teórica de la Materia Condensada, Universidad Autónoma de Madrid, E28049 Madrid, Spain}

 \affiliation{Condensed Matter Physics Center (IFIMAC), Universidad Autónoma de Madrid, E28049 Madrid, Spain}
\author{Donato Farina\, \orcidlink{0000-0002-7248-664X}}
\email[Email: ]{donato.farina@unina.it}
\affiliation{Physics Department E. Pancini, Università degli Studi di Napoli Federico II, Complesso Universitario Monte S. Angelo, Via Cintia, 21, Napoli, 80126, Italy}
\affiliation{Istituto Nazionale di Fisica Nucleare (INFN), Sezione di Napoli, Complesso Universitario Monte S. Angelo, Via Cintia, 21, Napoli, 80126, Italy.
}
\date{\today}
\begin{abstract}
We develop a first-principles thermodynamic framework for non-Hermitian dynamics based on a post-selected version of the fluctuation theorem. This allows us to identify a quantity that remains positive throughout the non-Hermitian evolution and can be interpreted as the entropy production of the post-selected dynamics. We relate this quantity to previously proposed notions of non-Hermitian entropy and derive an associated second law. Furthermore, we establish a connection with information-theoretic quantities, in particular the Petz--Rényi divergences, and leverage this connection to derive upper and lower bounds. Finally, we decompose the entropy production into incoherent and coherent contributions, identifying distinctive features of the coherent term in the vicinity of exceptional points. We illustrate our results using a paradigmatic model of non-Hermitian evolution based on a two-level system.
\end{abstract}
\maketitle

Non-Hermitian physics is emerging as a unifying framework across condensed-matter, quantum physics, atomic, molecular, and optical physics \cite{ashida2020non, ll76-j2l5}. It offers an effective formalism for describing loss, gain, and measurement back-action, thereby extending conventional quantum dynamics in a physically meaningful way.
It also unveils peculiar phenomena, such as the emergence of exceptional points,
where two or more eigenvalues coalesce together with their corresponding eigenvectors \cite{heiss2012physics, minganti2019quantum},
and parity-time symmetry breaking \cite{PhysRevLett.80.5243,  naghiloo2019quantum,di2025tomographic}. In condensed matter it has opened new avenues, including novel topological phases \cite{PhysRevX.9.041015, PhysRevX.8.031079}, non-reciprocal responses \cite{PhysRevResearch.2.013058, RevModPhys.93.015005}, the non-Hermitian skin effect \cite{PhysRevLett.121.086803}, as well as advantages in quantum sensing \cite{PhysRevLett.112.203901, hodaei2017enhanced, chen2017exceptional} stimulating recent debate \cite{g3n3-gh49}.
Despite the impressive advances in the field, fundamental questions remain: what impact will non-Hermitian physics have on other foundational areas  \cite{ll76-j2l5, roccati2026perspective} such as quantum thermodynamics?
In fact, despite recent progress \cite{SergiZloshchastiev2016, di2026local, tusun2026quantum, yang2026symmetry, lan2026kubo}, the thermodynamics of non-Hermitian systems has still to be fully understood, both at a fundamental level and in connection with distinctive non-Hermitian features such as the emergence of exceptional points.
Motivated by these open questions, we approach non-Hermitian thermodynamics from fluctuation theorems (FTs) \cite{campisi2011, Espositoreview, soret2022thermodynamic}, establishing a firm foundation for the field from first principles.
Our strategy is naturally grounded on the fact that both FTs and non-Hermitian dynamics are based on the physics of quantum trajectories: while FTs are expressed as relations between the probability of a trajectory and its suitably defined backward one \cite{manzano2022quantum}, non-Hermitian evolution can be engineered via post-selection.
Indeed, it naturally emerges from conditioning the evolution on the no-jump trajectory \cite{minganti2019quantum, albarelli2024pedagogical, wauters2026engineering, settimo2026quantum, settimo2026stochastic}.
Building on these premises, we combine fluctuation theorems for quantum trajectories with post-selection \cite{PhysRevResearch.7.013077, fiusa2025counting} to develop a new framework for characterizing irreversibility in non-Hermitian systems and to define an associated entropy production.
This quantity can be expressed in terms of information-theoretic measures, including Petz--Rényi divergences, in close analogy with standard measures of irreversibility in Markovian open quantum systems \cite{breuer2002theory}, as well as in terms of signal-to-noise ratios of suitably defined observables.
We then derive the corresponding form of the second law, providing a formal explanation of why the average entropy variation in non-Hermitian systems is not necessarily positive. We conclude by discussing genuinely quantum features revealed by decomposing the entropy production into coherent and incoherent contributions, with the coherent term arising only for non-normal Hamiltonians. This decomposition provides a diagnostic for identifying distinctive features of non-Hermitian dynamics, including the presence of exceptional points, as illustrated through a paradigmatic example.

{\it {Preliminaries.---}}
Consider a quantum channel $\Phi$ on a system S originating from the contact with an equilibrium environment B at inverse temperature $\beta$, with $H_B = \sum_{\mu} \epsilon_{\mu} \ket{\mu}\bra{\mu}_B$ the Hamiltonian of the thermal bath in its spectral decomposition.
The channel admits a Kraus representation in which each Kraus operator $M_{\mu \nu}$ expresses a  jump between two eigenvectors of the environmental Hamiltonian $\ket{\nu}_B, \ket{\mu}_B$\,\cite{breuer2002theory},
\begin{equation} \label{eq:channel}
\Phi(\rho) = \sum_{\mu, \nu} M_{\mu \nu} \rho \, M^\dagger_{\mu \nu}\,,
\end{equation}
where $\sum_{\mu \nu} M_{\mu \nu}^{\dag} M_{\mu \nu} = \mathbb{1} $ and $\rho$ denotes the initial state of S.
In this setting, under the assumption of microreversibility
of the system-environment evolution, we can establish a correspondence between the Kraus operators of the time reversed process $\tilde{M}_{\mu \nu}$ and those of the forward process\,\cite{manzano2015nonequilibrium, manzano2018quantum}
\begin{equation} \label{eq:revkraus}
    \tilde{M}_{\mu \nu} =   M_{\nu \mu}^{\dag} e^{ \beta \frac{\omega_{\mu \nu}}{2}},
\end{equation}
where $\omega_{\mu \nu} = \epsilon_{\mu} - \epsilon_{\nu} $ \footnote{Notice that, differently from \cite{manzano2018quantum}, we absorbed a system time-reversal operator $\Theta$ in the definition of the time-reversed process}.
Fluctuating heat and entropy variation can be quantified by using a two-point measurement (TPM) scheme~\cite{kurchan2001, talkner2016aspects, Gaspard}.
Given an initial density matrix
{in its spectral decomposition}
$\rho = \sum_m p_m \ket{m}\bra{m}$, the joint probability of the system-bath compound to be in an initial state $\ket{\nu}_B \ket{m}$ and a final state $\ket{\mu}_B \ket{\tilde{n}}$ is given by 
\begin{equation} \label{eq:probforw}
    P_{\nu\rightarrow \mu}(n,m) = | \bra{\tilde{n}}  M_{\mu \nu}  \ket{m}|^2 p_m.
\end{equation}
The probability in the backward process is similar, up to replacing the Kraus operators with their time-reversed counterparts $\tilde{M}_{\nu \mu}$ and introducing an initial preparation of the time-reversed process $\tilde{\rho}=  \sum_n \tilde{p}_n \ket{\tilde{n}}\bra{\tilde{n}}$,
\begin{equation} \label{eq:probback}
     \tilde{P}_{\mu \rightarrow \nu}(m,n) = | \bra{m}  \tilde{M}_{\nu \mu}  \ket{\tilde{n}}|^2 \tilde{p}_n.
\end{equation}
Note that, at this point of the discussion, $\tilde{\rho}$ is an arbitrary state, and different choices result in the derivation of fluctuation relations for several variants of the entropy production \cite{manzano2018quantum, esposito2010three}.
After defining {the stochastic system entropy change} $\Delta S_{nm} \coloneqq  \log p_m - \log \tilde{p}_n$, and considering the probability $P$ of obtaining a fixed variation $\Delta S_{nm} = \Delta S$,
by employing Eqs.\,\eqref{eq:revkraus} and \eqref{eq:probback}
we have that (see App. \ref{app:FTapp})
\begin{equation} \label{eq:flcutold}
    \frac{P_{\nu \rightarrow \mu}(\Delta S)}{\tilde{P}_{\mu \rightarrow \nu}(
    -\Delta S)} = e^{-\beta  \omega_{\nu \mu} + \Delta S }.
\end{equation}
This fluctuation theorem serves as a reference point for our extension to the non-Hermitian setting.

{\it Post-selection and fluctuation theorem.---}
The equation of motion described by a non-Hermitian Hamiltonian {$H_{\rm eff}$} in its non-trace preserving formulation is
\begin{eqnarray} \label{eq:nonHdyn}
    \dot{\Omega}(t)= - i (H_{\rm eff} (t) {\Omega}(t) - {\Omega}(t) H_{\rm eff}^\dag (t))\,,
\end{eqnarray}
where the system state ${\Omega}(t)$ is, in general, not normalized.
This can be obtained by post-selecting a single environmental history in which no jumps occurred \cite{minganti2019quantum}.
Indeed, consider the zero-temperature case in which a ground state of the environment is post-selected to be stationary; we then have
$\mu = \nu = 0$ in Eq. \eqref{eq:channel}, $U_{\rm eff}:=M_{ 00}$, obtaining
\begin{equation}
\label{evolved-nh}
    {{\Omega}(t)} = U_{\rm eff} (t)\rho U_{\rm eff}^{\dag}(t)\,,
\end{equation}
where ${\dot U_{\rm eff} (t) = -i H_{\rm eff} (t) U_{\rm eff} (t)}$ (having set $\hbar=1$).
In turn, for a time-independent $H_{\rm eff}$, the operator $U_{\rm eff}$ entering Eq.\,\eqref{eq:nonHdyn} reduces to $U_{\rm eff} = \exp{\big(-iH_{\rm eff} {t}\big)}$.
By doing the same post-selection choice in Eq.\,\eqref{eq:flcutold}, the fluctuation relation becomes\,
\begin{equation}
   \frac{P_{0 \rightarrow 0}(\Delta S)}{\tilde{P}_{0 \rightarrow 0}(
    -\Delta S)} = e^{\Delta S }.
\end{equation}
Although similar, this is not yet in the form of a FT, since the quantities at the left-hand side are not probabilities.
However, if we introduce proper normalization constants $\mathcal{N} ,\:\tilde{\mathcal{N}}$, together with probabilities $P(\Delta S), \: \tilde{P}(\Delta S)$ such that $P(\Delta S) \coloneqq \mathcal{N} P_{0 \rightarrow 0}(\Delta S)$,
 $\tilde{P}(\Delta S) \coloneqq \tilde{\mathcal{N}} \tilde{P}_{0 \rightarrow 0}(\Delta S)$ we obtain
\begin{equation} \label{eq:newfluct}
   \frac{P(\Delta S)}{\tilde{P}(
    -\Delta S)} = e^{\Delta S + \Xi},
\end{equation}
where
 $ \Xi \coloneqq  \log \mathcal{N} - \log \mathcal{\tilde{N}}$ is a genuine non-Hermitian correction.
The quantity appearing in the exponent on the right-hand side of Eq.~\eqref{eq:newfluct} can be interpreted as the stochastic entropy production associated with the non-Hermitian evolution; a more precise discussion is provided in the next section. 
It is convenient to define its average over all the post-selected trajectories, leading to the following definition:
 \begin{equation} \label{eq:avgfluct}
\Sigma \coloneqq \langle \Delta S \rangle + \Xi\,.
\end{equation}
%where the average is performed on all the possible realizations of the process.
We remark that all the results above are true for any choice of initial backward state $\tilde{\rho}$. From now on, we will make the choice $\tilde{\rho} = \rho(t) \coloneqq  \Omega(t)/ {\rm Tr}[\Omega(t)]$, the normalized final state.

{ \it Entropy production for non-Hermitian dynamics.---}
After some algebra on $\mathcal{N}, \tilde{\mathcal{N}}$ using the definitions \eqref{eq:probforw}, \eqref{eq:probback} and the property \eqref{eq:revkraus}, it is possible to express the genuine non-Hermitian correction $  \Xi$ as a function of ${Q \coloneqq U_{\rm eff}^{\dag} U_{\rm eff} \preceq \mathbb{1}}$ (see App. \ref{app:NHapp}),
\begin{equation}
\Xi =\log\!\left(\frac{\Tr(Q^2\rho)}{\Tr(Q\rho)^2}\right)
=\log\!\left(1+\xi^2\right),
\label{eq:Sigma_def}
\end{equation}
where
\begin{equation}
\xi^2\coloneqq \frac{\Var_{\rho}(Q)}{\langle Q\rangle_{\rho}^2}
\end{equation}
is the inverse square of the signal-to-noise ratio of the measurement of $Q$ on the initial preparation $\rho$.
Note that, since $\xi^2 $ is always positive, 
$\Xi$ is positive as well.
From Eq.\,\eqref{eq:newfluct} it immediately follows that
\begin{eqnarray}
    \langle e^{-\Delta S} \rangle = e^{ \Xi}.
\end{eqnarray}
As usual, the concavity of the logarithmic function allows us to obtain the following second law-like statement:
\begin{eqnarray} \label{eq:2law}
    \langle \Delta S \rangle \geq  -  \Xi,
\end{eqnarray}
that is, using the definition in Eq.\,\eqref{eq:avgfluct}, $\Sigma \geq 0$.
The equation above expresses a limitation on the entropy variation of the system during the non-Hermitian evolution.
If the Hamiltonian is Hermitian, $Q = U_{\rm eff}^{\dag} U_{\rm eff} = \mathbb{1}$, so that $ \Xi =0$ and we recover $ \langle \Delta S \rangle \geq 0$, that is compatible with the value $\langle \Delta  S \rangle =0$ for a unitary evolution.
Two features clarify the physical meaning of $\Sigma$. 
First, by being always positive during the non-Hermitian dynamics and being directly associated to a FT, it can be genuinely interpreted as an entropy production.
Notice, however, that this interpretation holds only if considering the post-selected scenario, while $\Sigma$ bears no straightforward connection with the entropy production of the underlying pre-selected physical model. This is clearly implied by the fact that the heat exchange contribution of Eq.~\eqref{eq:flcutold} is completely absent in Eqs. \eqref{eq:newfluct} and \eqref{eq:avgfluct}.
Second, $\Sigma$ is invariant under an overall rescaling of the non-Hermitian Hamiltonian 
$H_{\rm eff}(t) \mapsto H_{\rm eff}(t) + i \alpha \id$, $\alpha$ being an arbitrary real number. Such a rescaling only affects the normalization, which leaves $\Xi$ and $S$ unchanged. This distinguishes $\Sigma$  from the Sergi--Zloshchastiev non-Hermitian entropy correction $-k_B\ln\Tr\Omega(t)$~\cite{SergiZloshchastiev2016}, which is sensitive to a normalization gauge that is unobservable if we assume, as we did in our post-selection scheme, to forget the information about the original, pre-selected scenario. 

{ \it Information theoretic reformulation.---}
Our results also admit a direct information-theoretic reformulation. Defining the auxiliary state
\begin{align}
\sigma:=\frac{\rho^{1/2}Q\rho^{1/2}}{\Tr(Q\rho)},
\end{align}
the non-Hermitian correction can be written as (see App. \ref{app:NHapp})
\begin{align}
\Xi
=
\log\Tr(\sigma^2\rho^{-1})
=
D_2(\sigma\|\rho) ,
\label{eq:nH_entropy_prod_renyi_entropy}
\end{align}
namely, the Petz--Rényi divergence of order $2$, with $\rho^{-1}$ understood on $\supp(\rho)$. 
Although for a generic state $\sigma$ the quantity $D_2(\sigma\|\rho)$ is equal to $+\infty$ whenever $\supp(\sigma)\nsubseteq\supp(\rho)$, this situation does not arise here, since $\sigma$ is automatically non-zero only on $\supp(\rho)$
\footnote{Indeed, for any $v\in\ker(\rho)$ one has $\rho^{1/2}v=0$, and therefore ${\sigma v \sim \rho^{1/2}Q\rho^{1/2}v=0}$. Thus $\ker(\rho)\subseteq\ker(\sigma)$, which is equivalent to $\supp(\sigma)\subseteq\supp(\rho)$.}.
This representation allows us to determine upper and lower bounds on ${\Xi}$.
Indeed, using well-known properties of Petz--Rényi divergences, it can be shown that 
\begin{align} \label{eq:boundtriplo1}
\Xi 
& \le
D_{\max}(\sigma\|\rho)
= \log \frac{q_{{\rm max}}}{\Tr[Q \rho]}, 
\end{align}
where $D_{\max}(\sigma\|\rho)
:=
\log \inf
\left\{
 c>0:\sigma \preceq c\rho
\right\}$ denotes the max-relative entropy~\cite{datta2009min,gour2025resources},
and $q_{{\rm max}}$ is the maximum eigenvalue of $Q$. 
The result above gives a bound on the non-Hermitian correction in terms of the comparison between the trace loss of the state and the trace loss of its components with the slowest decay (those lying in the eigenspace relative to $q_{\max})$.
The average entropy variation can be characterized in terms of a relative entropy as well (see App. \ref{app:NHapp}). We have
\begin{align}
\langle \Delta S\rangle=-D(\sigma\|\rho),
\end{align}
with $D$ the usual relative entropy. Therefore Eq.~\eqref{eq:2law} can also be obtained directly as
\begin{align} \label{eq:entropy_prod_functional}
\langle \Delta S\rangle+\Xi
=
D_2(\sigma\|\rho)-D(\sigma\|\rho)\ge 0,
\end{align}
where the inequality follows from the monotonicity of $\alpha$-Petz--Rényi relative entropies with respect to $\alpha$~\cite{hiai2024log}. This does not only provide a derivation complementary to the FT, but also allows one to characterize the saturation condition for Eq.~\eqref{eq:2law}, namely
$D_2(\sigma\|\rho)=D(\sigma\|\rho)$ which, by strict monotonicity in respect to the Rényi order~\cite{hiai2024log}, is equivalent to $\sigma=\rho$. In turn, this implies
\begin{align}
\frac{\rho^{1/2}Q\rho^{1/2}}{\Tr(Q\rho)}=\rho
\end{align}
namely, $Q$ must be proportional to the identity on the support of the initial state. Equivalently, we have 
$
\Var_{\rho}(Q)=0,$
that implies $\Xi =0$.
Hence Eq.~\eqref{eq:2law} can be saturated only in the trivial case $\Xi=0$, which also implies $\langle \Delta S\rangle=0$. 
In particular, whenever the non-Hermitian correction is strictly positive, the bound in Eq.~\eqref{eq:avgfluct} is necessarily strict.
Note that Eq. \eqref{eq:entropy_prod_functional} can also be seen as a lower bound on $\Xi$ in terms of $\langle \Delta S\rangle$. %that has a well-defined physical meaning.

All the bounds derived so far explicitly take into account the initial state $\rho$; going back to the characterization \eqref{eq:Sigma_def} it is immediate to find a state-independent upper bound. By maximizing the variance at fixed mean this is achieved by rank-2 states with weights only on the maximum  and minimum eigenvalues of $Q$
\begin{equation}
\Xi \leq 
\log\!\left[
\frac{(q_{\max}+q_{\min})^2}{4q_{\max}q_{\min}}
\right].
\label{eq:Xi_max_main}
\end{equation}
Given its derivation, the bound is saturable, providing a recipe for a state preparation that maximizes the non-Hermitian correction $\Xi$.

{\it {Coherent and incoherent contributions.---}}
Similarly to other entropic and thermodynamic measures \cite{PhysRevLett.113.140401, PhysRevLett.125.180603}, it is possible to split the non-Hermitian entropy production in a coherent and an incoherent contribution.
{We start by separating a time-independent non-Hermitian Hamiltonian in its Hermitian and anti-Hermitian terms,}
\begin{equation}
H_{\rm eff}=H-iF,
\label{eq:Meff_def}
\end{equation}
with \(H\) and \(F\) Hermitian.
We define the incoherent contribution to the non-Hermitian correction as ${\Xi_{i}:=\Xi\big|_{H=0}}$.
This naturally leads to the introduction of a coherent contribution, defined as
\begin{equation}
\Xi_{c}=\Xi-\Xi_{i}\,.
\label{coh-EP}
\end{equation}
The rationale behind these definitions is as follows. 
Whenever $[H,F]=0$, the Hamiltonian $H$ does not contribute to the entropy \eqref{eq:Sigma_def}, making it natural to identify the incoherent component by simply setting $H=0$. As a consequence, the coherent part $\Xi_{c}$ is nonzero only when the Hermitian and anti-Hermitian sectors do not commute. We further note that 
\begin{equation}
[H_{\rm eff},H_{\rm eff}^\dagger]
=
2i[H,F],
\label{eq:Heff_non_normality_main}
\end{equation}
so \([H,F]\neq0\) is precisely the non-normality condition for
\(H_{\rm eff}\). 
Condition \eqref{eq:Heff_non_normality_main} is not merely a formal one: non-normality is an important prerequisite for the observation of exotic physical behavior, as it is a necessary condition for the emergence of exceptional points~\cite{ashida2020non}.
If the dynamics exhibits exceptional points, then, for a suitable choice of the initial state, they can be detected by examining the qualitative behavior of $\Xi_c$. In particular, $\Xi_c$ displays distinctive features such as polynomial corrections to the otherwise oscillatory or exponentially decaying behavior, as well as a high sensitivity to parameter variations. A specific example is discussed in the next section.
It is clear from the definition that $\Xi_c$ is not necessarily a decreasing, nor an increasing, function of time.
Indeed, we can make either $\Xi$ or $\Xi_i$ equal to $0$ by choosing as an initial state an eigenoperator of $Q$  or $e^{-2 Ft}$, respectively. These states have zero variance, so they produce a vanishing non-Hermitian correction or a vanishing incoherent part of it, according to Eq. \eqref{eq:Sigma_def}.
Generally, the sign of \(\Xi_c\) is fixed by how the coherent dynamics
changes the relative fluctuations of \(Q\) with respect to the incoherent
baseline: defining
\(\xi^2_i:={\operatorname{Var}_{\rho}(e^{-2 Ft})}/
{\langle e^{-2 Ft} \rangle_{\rho}^{2}}\),
one has
\begin{equation}
\Xi_c
=
\log\!\left(
\frac{1+\xi^2}{1+\xi^2_i}
\right).
\label{eq:Sigma_eps}
\end{equation}
Thus \(\Xi_c>0\) if and only if the coherent dynamics increases the relative
fluctuations above the dissipative baseline set by $F$, while \(\Xi_c \leq 0\) otherwise.

{\it A paradigmatic qubit example.---} 
To illustrate the preceding ideas, we consider the post-selected dynamics of a two-level system generated by the effective Hamiltonian
\begin{align}
H_{\mathrm{eff}}
&=
J\bigl(\ket f\bra e+\ket e\bra f\bigr)
+
\bigl(\Delta-i\gamma_e/2\bigr)\ket e\bra e ,
\label{eq:Heff_qubit}
\end{align}
where ${J,\Delta\in \mathds R}$ and ${\gamma_e>0}$.
This effective Hamiltonian appears, for instance, in the no-click trajectory of a three-level transmon circuit~\cite{naghiloo2019quantum}.  For this model, the operator $Q$ admits the Pauli decomposition 
\begin{align}
Q(t)
&=
A(t)\,\id+\mathbf B(t)\cdot\boldsymbol\sigma 
\label{eq:Q_pauli}
\end{align}
(see App.~\ref{app:effective_qubit_variance} for explicit expressions of the coefficients).
For an input qubit state
\(\rho=\frac12(\id+\mathbf r\cdot\boldsymbol\sigma)\), with 
\(|\mathbf r|\le1\), the mean and variance of \(Q\) are
\begin{align}
\Tr(Q\rho)
&=
A+\mathbf B\cdot\mathbf r,
\quad
\operatorname{Var}_{\rho}(Q)
=
|\mathbf B|^2-(\mathbf B\cdot\mathbf r)^2 ,
\label{eq:qubit_mean_var}
\end{align}
and hence
\begin{align}
\Xi
&=
\log\!\left[
1+
\frac{|\mathbf B|^2-(\mathbf B\cdot\mathbf r)^2}
{(A+\mathbf B\cdot\mathbf r)^2}
\right].
\label{eq:Xi_qubit}
\end{align}
\begin{comment}  Indeed,
\begin{equation}
{[F,Q]=\frac{i\gamma_e}{2}
\bigl(B_x\sigma_y-B_y\sigma_x\bigr)},
\label{eq:comm_fq}
\end{equation}
so \(Q\) and \(F\) commute only when \(B_x=B_y=0\).  Whenever
\((B_x,B_y)\neq(0,0)\), one may choose \(\rho\) as a \(Q\)-eigenstate that
is not an \(F\)-eigenstate; then \(\Xi=0\), while \(\Xi_i>0\), and
therefore \(\Xi_c<0\). 
\end{comment}
Exceptional points of \cref{eq:Heff_qubit} occur at $\Delta=0$, $|J|=\gamma_e/4$, and are here, for convenience, distinguished into a right EP ($J=+\gamma_e/4$) and a left EP ($J=-\gamma_e/4$).
At these points we have, respectively,
\begin{align}
&\mathbf B_{\rm EP}(t)= \mp\frac{\gamma_e t}{2}e^{-\gamma_e t/2}
\left(0,\,\gamma_e t/4, \,1\right),
\label{eq:AB_EP_qubit}
\end{align}
so the \(Q\)-eigenaxis lies in the \(yz\)-plane. 
At the right EP, \cref{eq:AB_EP_qubit} gives the Bloch axis relative to the unique eigenvector, that is
\begin{align}
\mathbf r_{\rm EP}^{(+)}
=
\frac{
\left(
0,
\gamma_e t/4,
1
\right)}
{\sqrt{1+(\gamma_e t/4)^2}} .
\label{eq:r_EP_plus_direction}
\end{align}
%For \(\gamma_e=1\) and \(t=3\), this gives
%\(\mathbf r_{\rm EP}^{(+)}=(0,0.6,0.8)\).

%
\begin{figure}[t]
  \centering
    \includegraphics[width=1\linewidth]{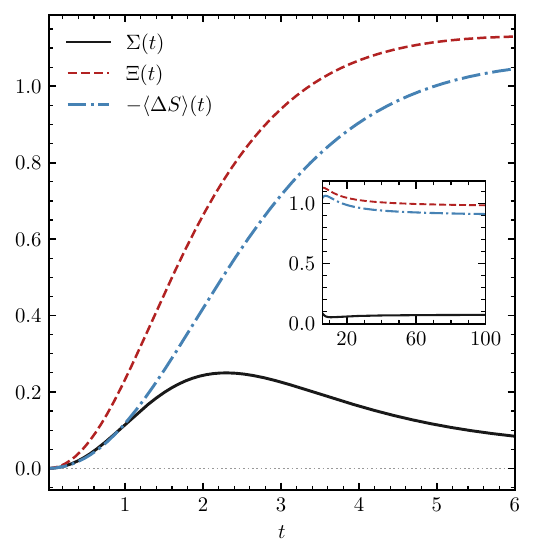}
  \caption{\justifying
  \(\Sigma(t)\), \(\Xi(t)\), and \(-\langle\Delta S\rangle(t)\) for the
  effective qubit model \cref{eq:Heff_qubit} at the right EP
  (\(\Delta=0\), \(J=\gamma_e/4\)), for the mixed input
  \(\rho=\frac12(\id+s\,\mathbf r_{\rm EP}^{(+)}\cdot\boldsymbol\sigma)\)
  with \(s=0.4\) and \(\gamma_e=1\). Inset: same quantities over the
  long-time range \(t\in[6,100]\).}
  \label{fig:sigma-deltaS-vs-t}
\end{figure}

To test Eq.\,\eqref{eq:avgfluct} we initialize the dynamics in a mixed state with Bloch vector $s\, \mathbf r _{\rm EP}^{(+)}$, where %$r _{\rm EP}^{(+)}$ the right EP and
$s \leq 1$, and study \(\Sigma(t)\) and \(\langle \Delta S\rangle(t)\) as functions of time. %postselected dynamics unfold, 

We observe that \(\Xi(t)\) and \(-\langle \Delta S\rangle(t)\) grow
in time as \(\rho(t)\) becomes increasingly distinguishable from the input,
while \(\Sigma(t)\) displays a mild non-monotonic transient before saturating (see \cref{fig:sigma-deltaS-vs-t}).

\begin{figure}[t]
  \centering
    \includegraphics[width=\linewidth]{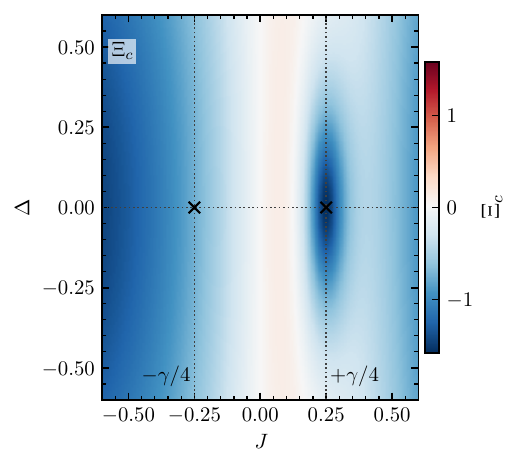}
  \caption{\justifying
  Coherent contribution \(\Xi_c\), Eq.~\eqref{eq:Sigma_eps}, for the effective qubit
  model \cref{eq:Heff_qubit}, using the fixed right-EP input
  \(\rho=\frac12(\id+\mathbf r_{\rm EP}^{(+)}\cdot\boldsymbol\sigma)\),
  with \(\mathbf r_{\rm EP}^{(+)}=(0,0.6,0.8)\), \(\gamma_e=1\), and \(t=3\).
  The exceptional points \(\Delta=0\),
  \(J=\pm\gamma_e/4\), are marked. The negative region near the right EP
  shows that coherent non-Hermitian mixing can reduce the fluctuation
  correction below the dissipative baseline, although the total correction
  remains nonnegative.}
  \label{fig:ep-density-fixed}
\end{figure}

To investigate the signatures of the EP in the coherent contribution $\Xi_c$, we compute it by subtracting the incoherent term $\Xi_i$ from $\Xi$. The incoherent contribution is obtained by setting $\Delta=J=0$ in Eq.~\eqref{eq:Heff_qubit}, that is, by replacing $H_{\rm eff}$ with ${F=\frac{\gamma_e}{2}\ket e\!\bra e=\frac{\gamma_e}{4}(\id+\sigma_z)}$.
We plot \(\Xi_c\) as a function of \(\Delta\) and \(J\), keeping the input \(\mathbf r_{\rm EP}^{(+)}\) and the time fixed (see \cref{fig:ep-density-fixed}).
The plot displays a region of strongly negative values centered exactly on $J= {\gamma}/{4}$, $\Delta =0$, signaling the presence of the right exceptional point.
This is the combination of two effects.
First, the input state is an eigenvector of $Q$, but not of $F$. As a consequence, the variance vanishes and $\xi=0$ in Eq.~\eqref{eq:Sigma_eps}, while the incoherent contribution $\xi_i$ remains finite, implying that $\Xi_c<0$.

Second, the defective eigenstructure of the exceptional point makes
the \(Q\)-eigenaxis rotate rapidly with \(J\), so an input locked to one EP
is maximally sensitive to coherent perturbations, governed by the non-commutativity described in Eq.~\eqref{eq:Heff_non_normality_main}.
As a result, near the right EP we see a drastic variation of the \(\Xi_c\) even for small perturbations of the parameters.
Notice that there are other points in the plot for which $\Xi_c \leq 0$; however, unlike the exceptional point, they are not associated with a sharp, localized variation. This suggests that what distinguishes an EP is not the value of $\Xi_c$ itself, but rather its enhanced susceptibility to parameter variations. 

{\it Conclusions.---}
We derived an entropy production associated with non-Hermitian dynamics directly from fluctuation theorems, thereby establishing a foundation for thermodynamic analyses in this setting.
Such an entropy production naturally decomposes into the average entropic change and a distinctive non-Hermitian contribution. It is also invariant under arbitrary overall rescalings of the non-Hermitian Hamiltonian and therefore captures a genuine property of the post-selected dynamics.

The non-Hermitian contribution can itself be decomposed into a purely dissipative part and a genuinely coherent contribution, the latter being activated by the noncommutativity of the Hermitian and anti-Hermitian components of the effective Hamiltonian.
We found that exceptional points, a direct manifestation of this noncommutativity, shape the coherent component of the non-Hermitian entropy production, thereby providing a thermodynamic signature of the post-selected exceptional-point structure.

Our work opens several directions for future investigation. We expect direct experimental tests of our results to be feasible with current platforms, including superconducting circuits \cite{naghiloo2019quantum} and quantum-optical setups \cite{ll76-j2l5, di2025tomographic}.

On the theoretical side, since our derivation applies equally to time-dependent non-Hermitian Hamiltonians \cite{figueira2006time}, it would be interesting to investigate phenomena that rely on time-dependent control, such as exceptional-point encircling \cite{doppler2016dynamically}, as well as intrinsically time-dependent settings, including non-Markovian dynamics.
In this direction, it would be particularly interesting to understand the interplay between post-selection and non-Markovianity, for instance by exploiting exact non-Markovian master equations \cite{d2025exact, pellitteri2026exact}.

%On a theoretical level, extending our formulation to generic non-trace-preserving quantum operations constitutes an interesting direction, as it is closely related to the thermodynamic role of generic quantum instruments.

{\it Acknowledgments.---}
V.C. and D.F. acknowledges financial support from PNRR MUR Project No. PE0000023-NQSTI.
F.E.Q.R. acknowledges financial support from the EU (ERC grant TIMELIGHT, GA 101115792) and MCIUN/AEI (PID2022-141036NA-I00 through MCIUN/AEI/10.13039/501100011033 and FSE+; RYC2021-031568-I; Programme for Units of Excellence in R\& D CEX2023-001316-M).

\bibliographystyle{apsrev4-2}
\bibliography{bibfile.bib}

\appendix
\begin{widetext}
    
\pagebreak
\section{Fluctuation theorem and entropy production}
\label{app:FTapp}
Eq. \eqref{eq:flcutold} can be obtained directly by defining $P_{\nu \rightarrow \mu}(\Delta S)$ and applying the symmetry \eqref{eq:revkraus}.
We have
\begin{align} \label{eq:ptransapp}
P_{\nu \rightarrow \mu}(\Delta S)
 \coloneqq \sum_{nm} P_{\mu \rightarrow \nu}(n,m) \delta(\Delta S - \Delta S_{nm})
 =\sum_{nm} | \bra{\tilde{n}}  M_{\mu \nu}  \ket{m}|^2 p_m \delta(\Delta S - \Delta S_{nm}).
\end{align}
We can now use the symmetry \eqref{eq:revkraus}
and the definition \eqref{eq:probback} in the formula for the forward transition probability:
\begin{align}
| \bra{\tilde{n}}  M_{\mu \nu}  \ket{m}|^2   = | \bra{m}  M_{\mu \nu}^\dagger  \ket{\tilde{n}}|^2
 = | \bra{m}  \tilde{M}_{\nu \mu}   \ket{\tilde{n}}|^2 e^{ -\beta \omega_{\nu \mu}}
= \tilde{P}_{\mu \rightarrow \nu}(m,n) e^{- \beta \omega_{\nu \mu}} \tilde{p}_n^{-1}, \notag
\end{align}
and after replacing the result inside Eq.\,\eqref{eq:ptransapp}, we obtain
\begin{equation} \label{eq:ptransapp1}
 P_{\nu \rightarrow \mu}(\Delta S) =
 \sum_{nm} \tilde{P}_{\mu \rightarrow \nu}(m,n) e^{- \beta \omega_{\nu \mu}} \frac{p_m}{\tilde{p}_n} \delta(\Delta S - \Delta S_{nm}).
\end{equation}
Inserting the definition of the fluctuating entropy variation $e^{\Delta S_{nm}} = p_m \tilde{p}_n^{-1}$ the equation above yields
\begin{align} \nonumber
 P_{\nu \rightarrow \mu}(\Delta S) & =
 \sum_{nm} \tilde{P}_{\mu \rightarrow \nu}(m,n) e^{- \beta \omega_{\nu \mu}} e^{\Delta S_{nm}} \delta(\Delta S - \Delta S_{nm}) \\
 &= \sum_{nm} \tilde{P}_{\mu \rightarrow \nu}(m,n) e^{- \beta \omega_{\nu \mu}} e^{\Delta S} \delta(\Delta S - \Delta S_{nm}). \label{eq:ptransapp2}
\end{align}

We introduce the transition probability at fixed entropy variation for the time-reversed process as
\begin{eqnarray} \nonumber
    \tilde{P}_{\nu \rightarrow \mu}(\Delta S)
    \coloneqq \sum_{nm} \tilde{P}_{\nu \rightarrow \mu}(n,m)
    \delta(\Delta S + \Delta S_{mn}),
\end{eqnarray}
where the plus sign and the change of indices in the $\delta$ function keep track of the fact that the time-reversed process starts from the distribution $\tilde{p}$, 
so the proper choice of initial stochastic entropy is $ - \Delta S_{mn} = \log \tilde{p}_m - \log p_n$.
We immediately see that
\begin{eqnarray} \nonumber
 \tilde{P}_{\mu \rightarrow \nu}(-\Delta S)
    \coloneqq \sum_{nm} \tilde{P}_{\mu \rightarrow \nu}(m,n)
    \delta(\Delta S - \Delta S_{nm}).  
\end{eqnarray}
that combined with Eq.\,\eqref{eq:ptransapp2} gives Eq.\,\eqref{eq:flcutold} of the main text.

\section{Explicit characterization of non-Hermitian entropy production}
\label{app:NHapp}

Starting from the definition of the normalization factors, we have
\begin{align} \label{eq:norm}
    \mathcal{N}^{-1} = \sum_{\Delta S} P_{0 \rightarrow 0}(\Delta S), \quad
      \tilde{\mathcal{N}}^{-1} = \sum_{\Delta S} \tilde{P}_{0 \rightarrow 0}(-\Delta S),
\end{align}
from which we obtain 
\begin{align}
\mathcal{N}^{-1} = \sum_{nm}
|\bra{\tilde{n}} U_{\rm eff} \ket{m}|^2 p_m
= \sum_{nm}  \bra{\tilde{n}} U_{\rm eff} \ket{m} \bra{m} U_{\rm eff}^{\dag} \ket{\tilde{n}} p_m = \Tr[ Q \rho],
 \\
\tilde{\mathcal{N}}^{-1} = \sum_{nm}  |\bra{m} \tilde{U}_{\rm eff} \ket{\tilde{n}}|^2 \tilde{p}_n = \sum_{nm}  \bra{m} U_{\rm eff}^{\dag} \ket{\tilde{n}} \bra{\tilde{n}} U_{\rm eff} \ket{m} \tilde{p}_n = \Tr[\tilde{Q}  \rho(t)], \label{eq:normapp}
\end{align}
where ${Q} \coloneqq 
U_{\rm eff}^{\dagger} U_{\rm eff}$, $\tilde{Q} \coloneqq U_{\rm eff} U_{\rm eff}^{\dagger}$ and $\rho(t)$ is the normalized evolved state introduced in the main text. In the fourth line we used $\tilde{U}_{\rm eff} = U_{\rm eff}^{\dagger}$, which follows from Eq.~\eqref{eq:revkraus} since $\omega_{00}=0$.
The last term of Eq.~\eqref{eq:normapp} can also be expressed as
\begin{equation}
  \Tr[\tilde{Q}  \rho(t)]
= \frac{\Tr[U_{\rm eff}^{\dag }\tilde{Q} U_{\rm eff} \rho]}{\Tr[Q \rho]}
= \frac{\Tr[Q^2 \rho]}{\Tr[Q \rho]},
\end{equation}
where we used the explicit form of the normalized final state of the non-Hermitian evolution,  $\rho(t) = U_{\rm eff}\rho U_{\rm eff}^{\dag}/\Tr[Q\rho]$, together with the cyclicity of the trace.
Combining the two results, we conclude
\begin{align} 
    &\Xi = \log \mathcal{N} -  \log \tilde{\mathcal{N}}
\\&=   -\log  \Tr[ Q \rho] +  \log \!\left(\frac{\Tr[Q^2 \rho]}{\Tr[Q \rho]}\right) = \log \!\left(\frac{\Tr[Q^2 \rho]}{\Tr[Q \rho]^2}\right), \nonumber
\end{align}
which is the result reported in Eq.~\eqref{eq:Sigma_def} of the main text.

We proceed by connecting the results above with the information-theoretic identities
\begin{align}
\langle \Delta S\rangle=-D(\sigma\|\rho),
\quad
\Xi=D_2(\sigma\|\rho),
\label{eq:app_goal_relent}
\end{align}
with
\begin{align}
\sigma:=\frac{\rho^{1/2}Q\rho^{1/2}}{\Tr(Q\rho)},
\label{eq:app_sigma_def}
\end{align}
where $D$ and and $D_2$ the Umegaki relative entropy and the Petz--Rényi relative entropy of order $2$, respectively.

The expectation value of the entropy variation in the forward, post-selected, process reads
\begin{align}
  & \sum_{\Delta S} P(\Delta S)
   \Delta S = 
\sum_{\Delta S} \mathcal{N}  P_{0 \rightarrow 0}(\Delta S)  \Delta S  =  \sum_{nm} P(n,m) \Delta S_{nm},
\end{align}
where we defined
\begin{eqnarray}
    P(n,m) \coloneqq \mathcal{N} P_{0 \rightarrow 0}(n,m).
\end{eqnarray}
We compute the two marginals and obtain 
\begin{align} \nonumber
    \sum_m  P(n,m) & = \mathcal{N} \sum_m 
    | \bra{\tilde{n}}  M_{0 0}  \ket{m}|^2 p_m = \mathcal{N} \sum_m \bra{\tilde{n}} U_{\rm eff} \ket{m} \bra{m} U_{\rm eff}^{\dag} \ket{\tilde{n}} p_m
   =   \frac{\bra{\tilde{n}} U_{\rm eff} \rho U_{\rm eff}^{\dag} \ket{\tilde{n}}}{\Tr[Q \rho]} = \tilde{p}_n,
\end{align}
\begin{align} \nonumber
    \sum_n P(n,m) & = \mathcal{N} \sum_n 
    | \bra{\tilde{n}}  M_{0 0}  \ket{m}|^2 p_m  = \mathcal{N} \sum_n \bra{\tilde{n}} U_{\rm eff} \ket{m} \bra{m} U_{\rm eff}^{\dag} \ket{\tilde{n}} p_m
   =  \frac{\Tr[Q \ket{m}\bra{m} ] }{\Tr[Q \rho]} p_m.
\end{align}
By replacing the marginals inside the formula for the average entropy we obtain
\begin{align}
\langle \Delta S\rangle
&=
\sum_{m,n}P(n,m)\log p_m
-
\sum_{m,n}P(n,m)\log \tilde{p}_n
=
\sum_{m}  \frac{\Tr[Q \ket{m}\bra{m} ] }{\Tr[Q \rho]} p_m \log p_m
-
\sum_{n} \tilde{p}_n \log \tilde{p}_n
\label{eq:app_avgS_split}
\end{align}
We now use that 
\begin{equation}
\sigma_m \coloneqq \sum_m \frac{\Tr[p_m^{1/2} Q \ket{m}\bra{m} p_m^{1/2}] \ket{m}\bra{m} }{\Tr[Q \rho]}
\end{equation}
commutes with $\sum_m \log p_m \ket{m}\bra{m}$
and we straightforwardly obtain
\begin{align}
\langle \Delta S\rangle
=
\Tr(\sigma\log\rho)-\sum_n \tilde{p}_n\log \tilde{p}_n
=\Tr(\sigma\log\rho)+S[\rho(t)].
\label{eq:app_avgS_intermediate}
\end{align}
To obtain a more compact form we show that the von Neumann entropies of $\sigma$ and $\rho(t)$ are the same. Indeed these two states can be expressed as
\begin{align} \label{eq:doubleA}
A A^{\dag}
=
\frac{U_{\rm eff}\rho U_{\rm eff}^\dagger}{\Tr(Q\rho )}
=
\rho(t),
\quad
A^{\dag} A
=
\frac{\rho ^{1/2}Q\rho ^{1/2}}{\Tr(Q\rho )}
=
\sigma,
\end{align}
where we introduced 
\begin{align}
A:=\frac{U_{\rm eff}\rho ^{1/2}}{\sqrt{\Tr(Q\rho )}}.
\end{align}
For any operator $A$, $AA^\dag$ and $A^\dag A$ have the same nonzero eigenvalues (with multiplicity): if $A^\dag A v=\lambda v$, $\lambda\neq0$, then $Av\neq0$ and $AA^\dag(Av)=\lambda(Av)$, and vice versa with $A\leftrightarrow A^\dag$. Only the zero eigenvalues, whose multiplicity depends on the dimensions of the domain and codomain, can differ.
Since $A A^{\dag}$ and $A^{\dag} A$ have the same nonzero spectrum, $\rho(t)$ and $\sigma$ have the same nonzero eigenvalues, and therefore
$
S(\rho(t))=S(\sigma) $, that substituted into \eqref{eq:app_avgS_intermediate} gives
\begin{align}
\langle \Delta S\rangle
=
\Tr(\sigma\log\rho )+S(\sigma)
=
\Tr(\sigma\log\rho )-\Tr(\sigma\log\sigma)
=
-\,D(\sigma\|\rho ),
\label{eq:app_avgS_relent}
\end{align}
which proves the first identity in \eqref{eq:app_goal_relent}. To prove the second identity in \eqref{eq:app_goal_relent} we start from Eq.\,\eqref{eq:Sigma_def}.
On the support of $\rho $, let $\rho ^{-1}$ denote the inverse of $\rho $ (equivalently, the Moore--Penrose pseudoinverse on the full Hilbert space).
From \eqref{eq:app_sigma_def},
\begin{align}
\sigma^2
=
\frac{\rho ^{1/2}Q\rho Q\rho ^{1/2}}{\Tr(Q\rho )^2},
\end{align}
and therefore
\begin{align}
\Tr(\sigma^2\rho ^{-1})
&=
\frac{1}{\Tr(Q\rho )^2}
\Tr\!\bigl(\rho ^{1/2}Q\rho Q\rho ^{1/2}\rho ^{-1}\bigr)
\nonumber\\
&=
\frac{1}{\Tr(Q\rho )^2}
\Tr(Q\rho Q)
=
\frac{\Tr(Q^2\rho )}{\Tr(Q\rho )^2},
\label{eq:app_sigma2_identity}
\end{align}
where in the last step we used the cyclicity of the trace. Hence
\begin{align}
\Xi
=
\log\Tr(\sigma^2\rho ^{-1})
=
D_2(\sigma\|\rho ),
\label{eq:app_Xi_D2}
\end{align}
which is the desired relation.

\section{Effective qubit Hamiltonian and variance formulas}
\label{app:effective_qubit_variance}

We collect here the elementary qubit identities used in the main text.  The effective two-level non-Hermitian Hamiltonian is
\begin{equation}
H_{\mathrm{eff}}
=
J\bigl(\ket f\!\bra e+\ket e\!\bra f\bigr)
+
\bigl(\Delta-i\gamma_e/2\bigr)\ket e\!\bra e ,
\label{eq:Heff_qubit_appendix}
\end{equation}
with \(J,\Delta\in\mathbb R\) and \(\gamma_e>0\).  In the ordered basis
\(\{\ket e,\ket f\}\) we have
\begin{equation}
H_{\mathrm{eff}}
=
\begin{pmatrix}
\Delta-i\gamma_e/2 & J\\
J & 0
\end{pmatrix},
\label{eq:Heff_qubit_matrix_appendix}
\end{equation}
that has eigenvalues 
\begin{equation}
\varepsilon_\pm
=
\frac{\Delta-i\gamma_e/2}{2}
\pm
\sqrt{
J^2+\frac14\bigl(\Delta-i\gamma_e/2\bigr)^2
}.
\label{eq:qubit_eigenvalues_appendix}
\end{equation}
Thus the two eigenvalues coalesce when the discriminant vanishes:
\begin{equation}
J^2+\frac14\bigl(\Delta-i\gamma_e/2\bigr)^2=0.
\label{eq:qubit_EP_discriminant_appendix}
\end{equation}
Since \(J,\Delta,\gamma_e\) are real and \(\gamma_e>0\), the condition on the imaginary part
forces \(\Delta=0\).  The real part then gives $J=\pm\frac{\gamma_e}{4}$.
At these points the matrix is not proportional to the identity and has a single %only
eigenvector with geometric multiplicity equal to 1, the degeneracy is therefore exceptional.
To find the explicit form of the operator $Q$ we work in the Pauli representation
\begin{equation}
\mathbbm 1=\ket e\!\bra e+\ket f\!\bra f,
\quad
\sigma_z=\ket e\!\bra e-\ket f\!\bra f,
\quad
\sigma_x=\ket e\!\bra f+\ket f\!\bra e .
\label{eq:Pauli_basis_qubit_appendix}
\end{equation}
The Hamiltonian can be written as
\begin{equation}
H_{\mathrm{eff}}
=
\zeta_0\mathbbm 1+\mathbf h\cdot\boldsymbol\sigma,
\label{eq:Heff_Pauli_appendix}
\end{equation}
where
\begin{equation}
\zeta_0
=
\frac{\Delta-i\gamma_e/2}{2},
\quad
\mathbf h
=
\left(
J,\,
0,\,
\frac{\Delta-i\gamma_e/2}{2}
\right).
\label{eq:zeta_h_defs_appendix}
\end{equation}
For ease of notation, we also introduce the following quantity, that will appear in further calculations:
\begin{equation}
\lambda \coloneqq \sqrt{\mathbf h\cdot\mathbf h}
=
\sqrt{J^2+\frac14\bigl(\Delta-i\gamma_e/2\bigr)^2}.
\end{equation}
We can now compute the exponential of the non-Hermitian Hamiltonian to obtain $U_{\rm eff}$ and $U_{\rm eff}^{\dag}$:
\begin{align}
U_{\rm eff}(t)
=
e^{-iH_{\mathrm{eff}}t}
=
e^{-i\zeta_0t}
e^{-it\,\mathbf h\cdot\boldsymbol\sigma}
=
e^{-i\zeta_0t}
\left[
c\,\mathbbm 1
-
i \frac{s}{\lambda} \,\mathbf h\cdot\boldsymbol\sigma
\right],
\label{eq:M_explicit_appendix}
\end{align}
where we introduced 
$
c:=\cos(\lambda t)$
and $s:=\sin(\lambda t)$.
Furthermore,
$
|e^{-i\zeta_0t}|^2=e^{-\gamma_e t/2}$
therefore we have 
\begin{align}
&Q(t)=
U_{\rm eff}^{\dag }(t)U_{\rm eff}(t)
= e^{-\gamma_e t/2}
\Bigl[
|c|^2\mathbbm 1
-i c^*\frac{s}{\lambda}\,\mathbf h\cdot\boldsymbol\sigma
+i\frac{s^*}{\lambda^*}\,\mathbf h^*\cdot\boldsymbol\sigma
+
|\frac{s}{\lambda}|^2
(\mathbf h^*\cdot\boldsymbol\sigma)
(\mathbf h\cdot\boldsymbol\sigma)
\Bigr].
\label{eq:Q_intermediate_appendix}
\end{align}
Our goal is to express $Q(t)$ as a linear combination of Pauli matrices, i.e. in the form $Q(t)
=
A(t)\mathbbm 1+\mathbf B(t)\cdot\boldsymbol\sigma $. We can use the following identity in Eq. \eqref{eq:Q_intermediate_appendix}
\begin{align}
(\mathbf h^*\cdot\boldsymbol\sigma)(\mathbf h\cdot\boldsymbol\sigma)
=
(\mathbf h^*\cdot\mathbf h)\mathbbm 1
+
i(\mathbf h^*\times\mathbf h)\cdot\boldsymbol\sigma ,
\label{eq:hstar_h_Pauli_product_appendix}
\end{align}
and finally obtain
\begin{align}
A(t)&=e^{-\gamma_et/2}\Big[|c|^2+|\beta|^2(\mathbf h^*\cdot\mathbf h)\Big]
=e^{-\gamma_et/2}\left(|c|^2+\frac{J^2+\Delta^2/4+\gamma_e^2/16}{|\lambda|^2}|s|^2\right),
\label{eq:A_explicit_appendix}\\
\mathbf B(t)&=e^{-\gamma_et/2}\Big[i(u\mathbf h^*-u^*\mathbf h)+i|\beta|^2(\mathbf h^*\times\mathbf h)\Big],
\label{eq:B_vector_general_appendix}
\end{align}
where we introduced $u:=\frac{s^*c}{\lambda^*}$, for ease of notation.
Since ${\mathbf h^*\times\mathbf h=(0,\,iJ\gamma_e/2,\,0)}$ 
the Bloch
vector of $Q(t)$ has components equal to 
\begin{equation}
B_x=-2Je^{-\gamma_et/2}u_I,
\qquad
B_y=-\frac{J\gamma_e}{2}e^{-\gamma_et/2}\frac{|s|^2}{|\lambda|^2},
\qquad
B_z=2e^{-\gamma_et/2}\Big[-\tfrac{\Delta}{2}u_I-\tfrac{\gamma_e}{4}u_R\Big],
\label{eq:B_components_appendix}
\end{equation}
where $u_I$ and $u_R$ are the imaginary and the real part of $u$.
 At the exceptional points, $\Delta=0$, $J=\pm\gamma_e/4$, one has $\lambda\to 0$,
 that implies $s \to 0$, so that 
$u=s^*c/\lambda^*$ and $|s|^2/|\lambda|^2$ becomes ratios of infinitely small quantities.
By Taylor expansion we obtain
\begin{equation}
u=\frac{s^*c}{\lambda^*}\ \xrightarrow{\ \lambda\to0\ }\ t
\quad\Longrightarrow\quad
u_R\to t,\qquad u_I\to0,
\label{eq:u_EP_limit}
\end{equation}
\begin{equation}
|s|^2=|\lambda|^2t^2+O(|\lambda|^4)
\quad\Longrightarrow\quad
\frac{|s|^2}{|\lambda|^2}\ \xrightarrow{\ \lambda\to0\ }\ t^2 .
\label{eq:s_over_lambda_EP_limit}
\end{equation}
Substituting \eqref{eq:u_EP_limit}--\eqref{eq:s_over_lambda_EP_limit} together with
$\Delta=0$ into \cref{eq:B_components_appendix} gives $B_x\to0$ (since $u_I\to0$)
and
\begin{equation}
\mathbf B_{\rm EP}(t)=
\left(0,\,-\frac{J\gamma_e}{2}e^{-\gamma_et/2}t^2,\,
-\frac{\gamma_et}{2}e^{-\gamma_et/2}\right),
\label{eq:AB_EP_qubit_derivation}
\end{equation}
recovering \cref{eq:AB_EP_qubit} in the main text.

\end{widetext}

\end{document}